\documentclass{aastex}
\usepackage{spr-astr-addons}
\usepackage{url}

\RequirePackage{color}

\begin{document}

\title{The mode switching of PSR B2020+28}
\shorttitle{The mode switching of PSR B2020+28}
\shortauthors{Z. G. Wen et al.}

\author{Z. G. Wen\altaffilmark{1,2}} \and \author{N. Wang\altaffilmark{1,3}} \and \author{W. M. Yan\altaffilmark{1,3}} \and \author{J. P. Yuan\altaffilmark{1,3}} \and \author{Z. Y. Liu\altaffilmark{1}} \and \author{M. Z. Chen\altaffilmark{1}} \and \author{J. L. Chen\altaffilmark{1,4}}
\and
\email{na.wang@xao.ac.cn}
\altaffiltext{1}{Xinjiang Astronomical Observatory (XAO), Chinese Academy of Sciences, Urumqi 830011, China.\\ {\it na.wang@xao.ac.cn}}
\altaffiltext{2}{University of Chinese Academy of Sciences, 19A Yuquan road, Beijing, 100049, China.}
\altaffiltext{3}{Key laboratory of Radio Astronomy, CAS, Nanjing, 210008, China.}
\altaffiltext{4}{Department of Physics \& Electronic Engineering, Yuncheng University, 044000, Yuncheng, Shanxi, China}

\begin{abstract}
This paper reports on polarimetric radiation properties based on the switching modes of normal PSR B2020+28 by analysing the data acquired from the Nanshan 25-m radio telescope at 1556 MHz. With nearly 8 hours quasi-continuous observation, the data presented some striking and updated phenomena. The change of relative intensity between the leading and trailing components is the predominant feature of mode switching. The intensity ratio between the leading and trailing components are measured for the individual profiles averaged over 30 seconds. It is found that there is an excess of high ratios over the normal distribution, which indicates that two modes exist in the pulsar. The distribution of abnormal mode has a narrower width indicating that the abnormal mode is more stable than the normal mode. A total of 76 mode switching events are detected in our data. It spends 89\% in the normal mode and 11\% in the abnormal mode. The intrinsic distributions of mode timescales are constrained with power-law distributions. The significant difference in the index of the duration distribution between normal and abnormal modes possibly indicates that the timescale for the abnormal mode to get stable is shorter than that for the normal mode. The frequent switching between both modes may indicate that the oscillations between different magnetospheric states are rapid.

\end{abstract}


\section{Introduction}

Pulsars are rapidly rotating, extremely dense neutron stars that emit radio frequency electromagnetic radiation from regions above their magnetic polar caps. As a pulsar rotates, the radiation originating from these regions sweeps through space much like a beam of a lighthouse. It sweeps across our line of sight once per rotation, hence we can receive the pulse signal. The pulsar presents a regularly spaced pulses with a repetition respectively stable period (hereafter $P$). In this case, the rotation frequency $\nu=1/P$. The individual pulses are very weak and vary with time significantly. In order to produce a stable profile, it requires the coherent addition of many hundreds or even thousands of pulses together. However, some pulsars show two or more patterns of averaged pulse profiles, which is called mode switching (or mode changing) phenomenon. Since the discovery of mode switching in PSR B1237+25 \citep{Backer+1970}, this phenomenon has been seen in a few dozens of pulsars. However, the integrated profiles of these pulsars are quite complex, e.g. PSR B1237+25 has 5 components \citep{Backer+1970,Rankin+1986}, PSR B0329+54 has triple components \citep{Rankin+1986,Chen+etal+2011}. The recent observations of PSRs B0723+26 \citep{Sobey+etal+2015} and B0943+10 \citep{Bilous+etal+2014} with LOFAR show two relatively stable states (bright and quiet modes). The forced Markov process was recently analysed by \citet{Cordes+2013} to model state changing pulsars.

The radio emission from pulsars is highly polarized \citep{Lyne+Smith+1968}. The mean pulse profile and the polarization measurements can yield a wealth of information, not only about the emission process itself, including the pulse emission mechanism, the beaming of pulsar radiation and the geometry of the system, but also about the medium through which it propagates. The integrated pulse profiles can be described with core/double-cone geometric model \citep{Rankin+1983}. The observed polarization position angle (hereafter $PA$) variations can be approximately described by the Rotating Vector Model (RVM) \citep{Radhakrishnan+Cooke+1969}. This model results from the idea that the radiation is polarized in the plane of curvature of field lines emanating from a magnetic pole on the star \citep{Komesaroff+1970}. For a simple dipole field, the observed PA variation is then determined by the projected direction of the magnetic axis as the star rotates. The rapid swings often observed near the profile midpoint imply that magnetic axis is nearly aligned with the observer's line of sight at that profile phase. The observed PA swings are various. \citet{Yan+etal+2011} presented polarization profiles of 20 pulsars with smooth, continuous and discontinuous PA variation. Discontinuities of approximately $90^\circ$ are often observed \citep{Manchester+etal+1975,Backer+Rankin+1980,Stinebring+etal+1984,Han+etal+2009}, and these are interpreted as resulting from overlapping emission from orthogonally polarized emission modes \citep{McKinnon+Stinebring+1998,Gangadhara+1997}. Such orthogonal polarization modes may have resulted from the radiation emitted by positrons and electrons while moving along the curved magnetic field lines \citep{Gangadhara+1997}, or generated as the wave propagates through the pulsar magnetosphere \citep{Petrova+2001}. The circular polarization is usually relatively weaker than the linear polarization. It is most often associated with the central or core component of the profile, often with a sense reversal near the profile mid-point \citep{Rankin+1983}.

PSR B2020+28 is a relatively strong pulsar with flux density of 38 mJy at 1400 MHz. The profile of this pulsar is relatively simple, having only two resolved components separated by 0.027 in pulse phase. The period P is 0.3434 seconds, and its first derivative is $\rm{1.89\times10^{-15}\, s\,s^{-1}}$. Correspondingly, its characteristic age ($\rm{\tau=2.87\times10^6\, years}$) is no more than the pulsar median. It was discovered with the east-west arm of the Bologna Cross telescope at 408 MHz \citep{Bonsignori-Facondi+etal+1973}. Polarimetric observations were obtained with the 76-m Lovell telescope at Jodrell Bank at radio frequencies centred around 230, 400, 600, 920, 1400, 1600 MHz, which gives the $W_{10}$, $W_{50}$, $W_e$, $L$, $|V|$, $V$ and polarization profiles respectively \citep{Gould+Lyne+1998}. There exist two orthogonal modes of polarization that have position angles separated by $90^\circ$ and opposite senses of circular polarization \citep{Cordes+etal+1978,Stinebring+etal+1984}. This pulsar also shows significant pattern change and is complicated by interstellar scintillation ~\citep{Wang+etal+2001}. However, the profiles of normal and abnormal modes have not been published so far. The radiation of various properties from pulsars are explained most naturally by a simple picture in which the observed radiation is produced by the acceleration of charged particles streaming outward along open filed lines above the poles of an essentially dipolar magnetic field.

In this paper we show a detailed investigation of the emission behavior of PSR B2020+28. In section 2, we describes the observing system and our observations. Section 3 presents data analysis and results. The implications of the results and conclusions are discussed in section 4.

\section{Observations}
\label{sect:Obs}

The observations of PSR B2020+28 were carried out using the Nanshan 25-m radio telescope on 2012 June 12. A dual-channel cryogenic receiver was used to receive orthogonal circular polarizations at a centre radio frequency of 1556 MHz, and with total bandwidth of 512 MHz. The system temperature, $T_{sys} = T_{rec} + T_{sky} + T_{spi}$ (in which $T_{rec}$, $T_{spi}$ and $T_{sky}$ are the receiver, spillover and the sky noise temperature, respectively), is approximately 32 K. An ortho-mode transducer (OMT) was used to resolve the electromagnetic wave into left-handed circular ($L$) and right-handed circular ($R$) basis modes. To determine the relative gain of the two polarization channels and the phase between them, a calibration signal is injected at an angle of $45^\circ$ to the feed probes. The two independent polarizations were then amplified and down-converted to an intermediate frequency in the range of 0$-$512 MHz with a local oscillator at 1300 MHz. The band-limited signals were fed to an updated back-end signal processing system, the third-generation Digital Filterbank System (DFB3), since 2010. After conversions from analogue voltages to digital signals at the Nyquist rate with 9-bit sampling, the DFB3 uses field-programmable gate array (FPGA) processors to produce a maximum of 8192 polyphase filterbank frequency channels and averaged pulse profiles. In order to obtain enough signal-to-noise ratio (S/N), about 90 pulse periods (~30 s) were averaged and a time resolution of 512 bins per pulse period was used in our observations. The data lasting nearly 8 hours from 1024 frequency channels were then recorded for off-line processing, which contain 75400 single pulses.

The data were calibrated carefully. The averaged pulses were obtained by de-dispersing the data at a dispersion measure (DM) of $24.6\ \rm{pc\ cm^{-3}}$. Four Stokes parameters were recorded and have been corrected for dispersion, interstellar Faraday rotation and various instrumental polarization effects following the method described by \cite{Yan+etal+2011}. The data are analysed offline using the PSRCHIVE package \citep{Hotan+etal+2004} and corrected for parallactic angle and the orientation of the feed. The four Stokes parameters are accessible in function (1),
\begin{equation}
\centering
  \left[\begin{array}{c}I \\
                         Q \\
                         U \\
                         V \\
        \end{array}\right]=\left[\begin{array}{c}|L|^2+|R|^2 \\
                                                  2Re(L*R)    \\
                                                  2Im(L*m)    \\
                                                  |R|^2-|L|^2   \\
                           \end{array}\right]
\label{eq:LebsequeI}
\end{equation}
where * indicates a complex conjugate \citep{Lorimer+2005}.

\section{Data reduction and analysis}

The structure and brightness of individual pulses are observed to vary significantly, but the average of many hundreds of individual pulses is usually stable, leading to a characteristic profile that is often unique to pulsar. However, some pulsars exhibits two or more discrete and well-defined pulse profile morphologies, and they switches abruptly, which is known as mode-switching. The change of relative intensity between the different components is the predominant feature of mode switching \citep{Chen+etal+2011}.

\begin{figure}[h]
   \centering
   \includegraphics[width=7.8cm, angle=0]{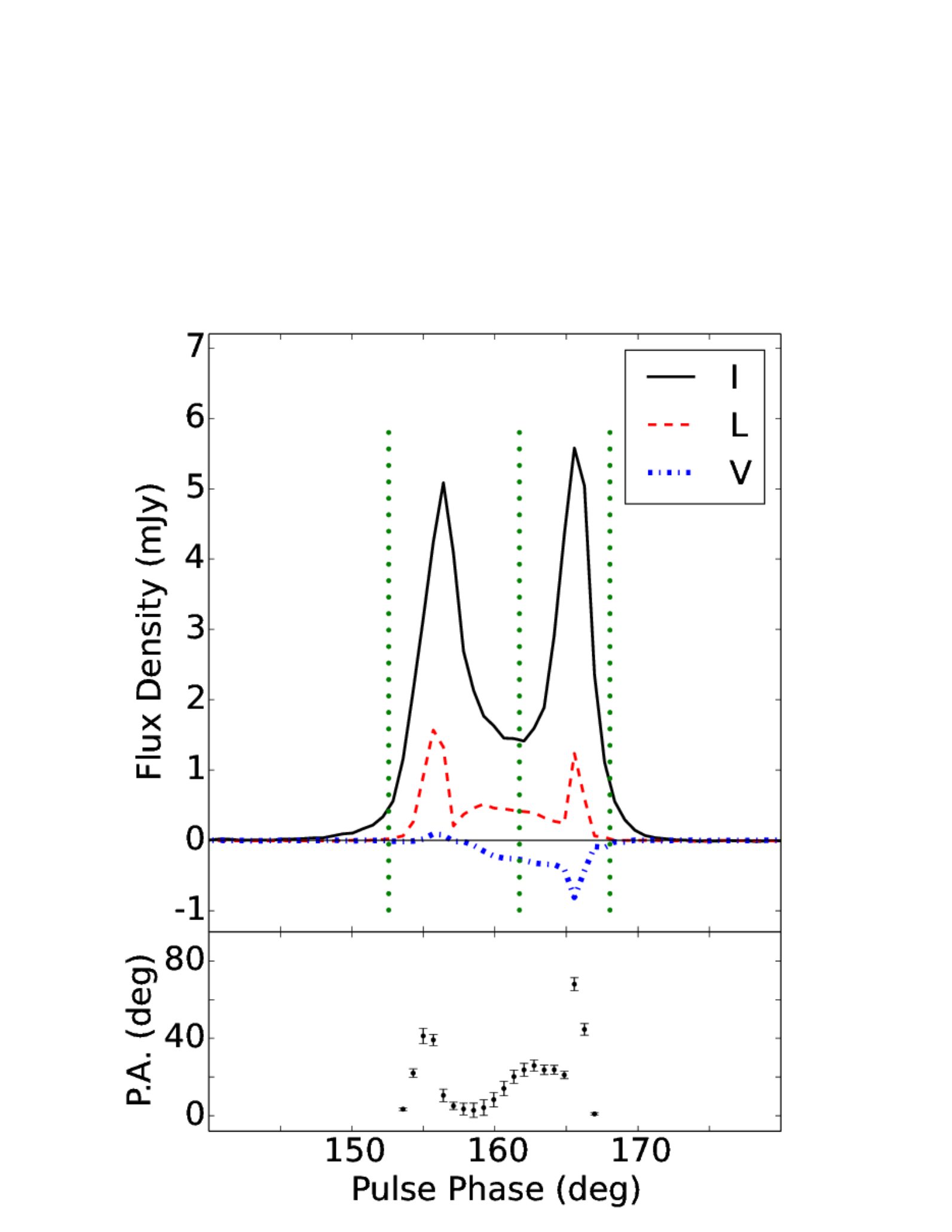}
   \caption{Polarization waveform of PSR B2020+28. The uniformly weighted values of the Stokes parameters are averaged over nearly 8 hours, with total intensity $I$, circular polarization $V$, and the linear polarization $L = (Q^2 + U^2)^{1/2}$ (where $Q$ and $U$ are the linear Stokes parameters), are displayed in solid, dot-dashed and dashed lines respectively.
%Stokes parameters $I$ (total intensity), $V$ (circular polarization), and the quantity $L = (Q^2 + U^2)^{1/2}$ (linear polarization), where $Q$ and $U$ are the linear Stokes parameters, are displayed in solid, dot-dashed and dashed lines respectively. 
The boundaries are plotted with dotted lines to distinguish leading and trailing components. The position angle as a function of pulse phase, $\chi =1/2\tan^{-1}(U/Q)$, is given in the lower panel.}
   \label{waveform}
\end{figure}

At 1.5 GHz, the average pulse profile of PSR B2020+28 is relatively simple, having double well resolved main emission components (leading and trailing components), which is common to many pulsars. They are connected by a pronounced bridge emission. The outer boundaries of main components are confined at 10\% of pulse peak of the average profile, while the inner boundary is defined as the minimal intensity between them. The examples of average pulse profile (solid line) and phase boundaries (vertical dotted lines) are shown in Fig. \ref{waveform}. The polarization characteristics of the mean pulse profile provide a framework for understanding the emission processes in pulsars. The averages of linear polarization (dashed line) is a maximum in the leading component at 30\% compared with 22\% in the trailing component. Two components are depolarized relative to low-frequency observations, but the effect is most noticeable in the leading component of the pulse \citep{Cordes+etal+1978,Stinebring+etal+1984}. The averages circular waveform (dash-dotted line) is almost negligible in the leading component and a sense reversal slightly trails the leading nulls in the linear polarization. The extrema of the circular polarization occur under the linear maximum in the trailing component. The position angle rotation (lower panel) in the saddle region shows a S-shaped sweep with total swing only about 30 degrees. Whereas, two peaks in the leading and trailing regions show up.

\begin{figure}[h]
   \centering
   \includegraphics[width=8.2cm, angle=0]{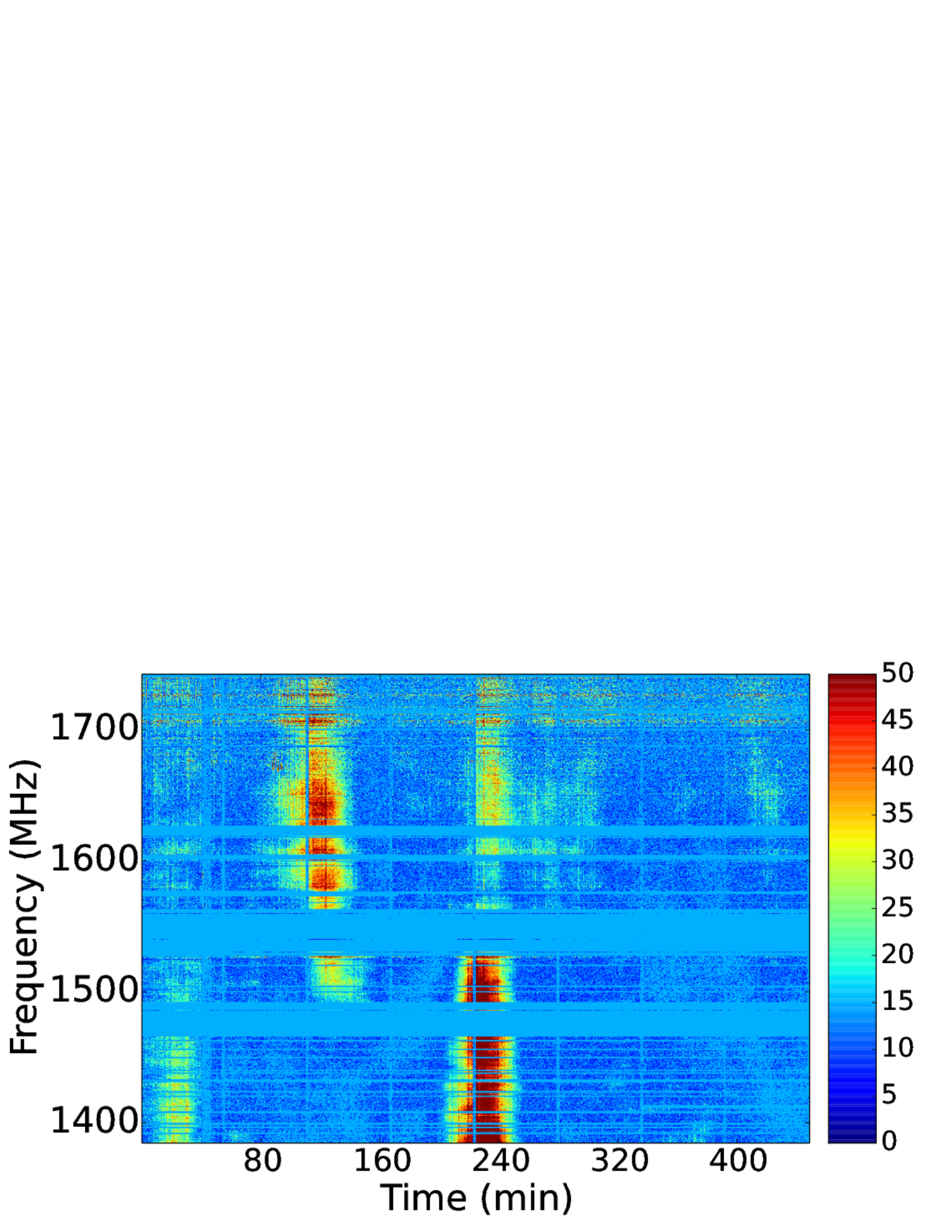}
   \caption{The dynamic spectrum of scintillations for PSR B2020+28. Here the measured signal to noise of the pulsar signal is plotted as a function of both time and frequency. The horizontal and vertical stripes in the dynamic spectrum are where data were affected by RFI.}
    \label{Dynamic_spectrum}
\end{figure}

The dynamic spectrum is shown in Fig. \ref{Dynamic_spectrum}. It seems like that the intensity fluctuations with time and frequency are very similar to the case of the observations by \citet{Wang+etal+2001}. The scintles (the red patches) have a characteristic timescale (the scintillation timescale) of three thousand seconds. The horizontal stripes in the dynamic spectrum are because the bad frequency channels are affected by Radio Frequency Interference (RFI). At the edges of the observing frequency, there is no useful data as the bandpass rolls off. The scintillation bandwidth (the characteristic frequency scale of the scintiles) is clearly less than the total bandwidth of the observation. The effects of such large fluctuations can be addressed by subtracting a running mean of length one fifth of the scintillation timescale.
%can be addressed by averaging each sub-integration in blocks of around a quarter of the scintillation timescale.

\begin{figure*}
   \centering
   \includegraphics[width=8.0cm, angle=0]{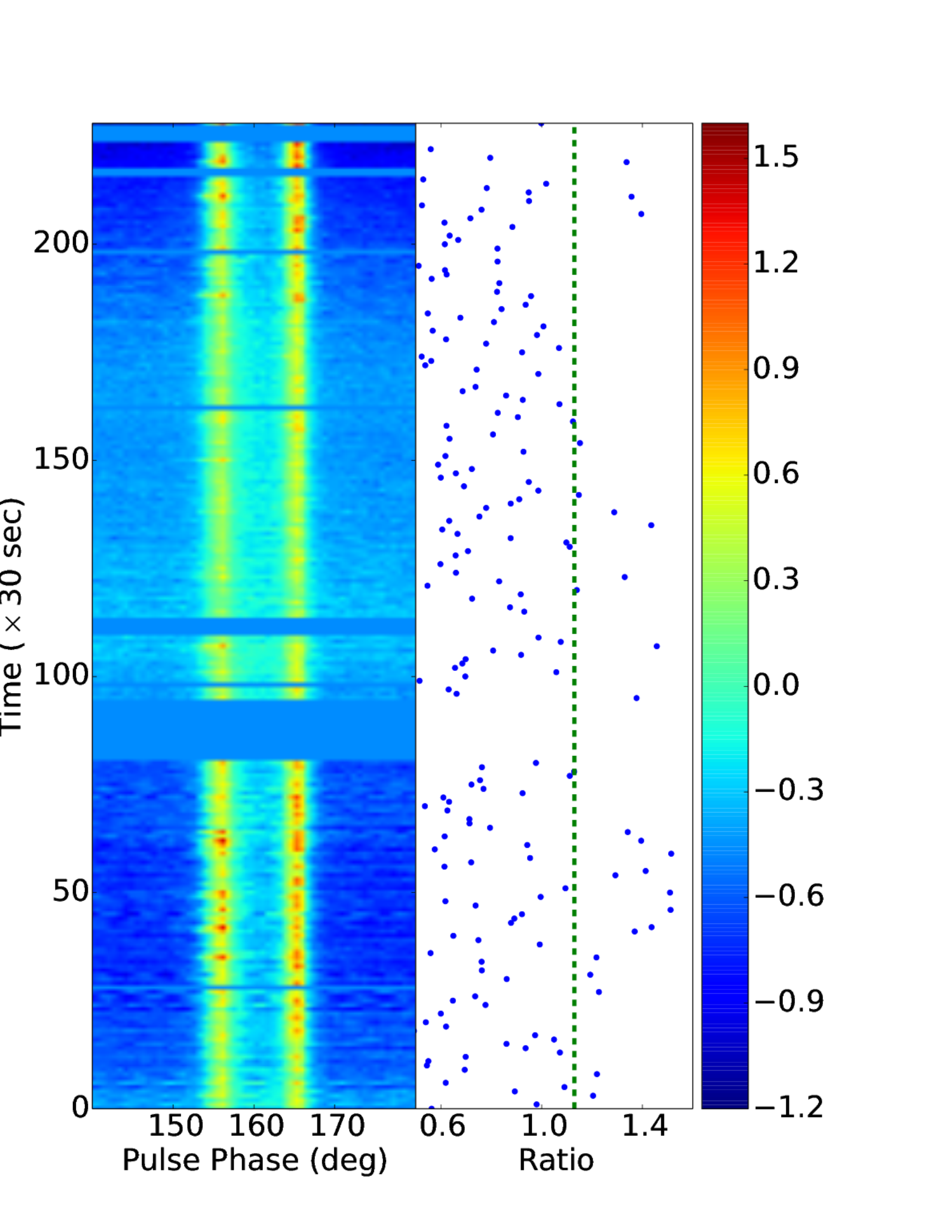}
   \includegraphics[width=8.0cm, angle=0]{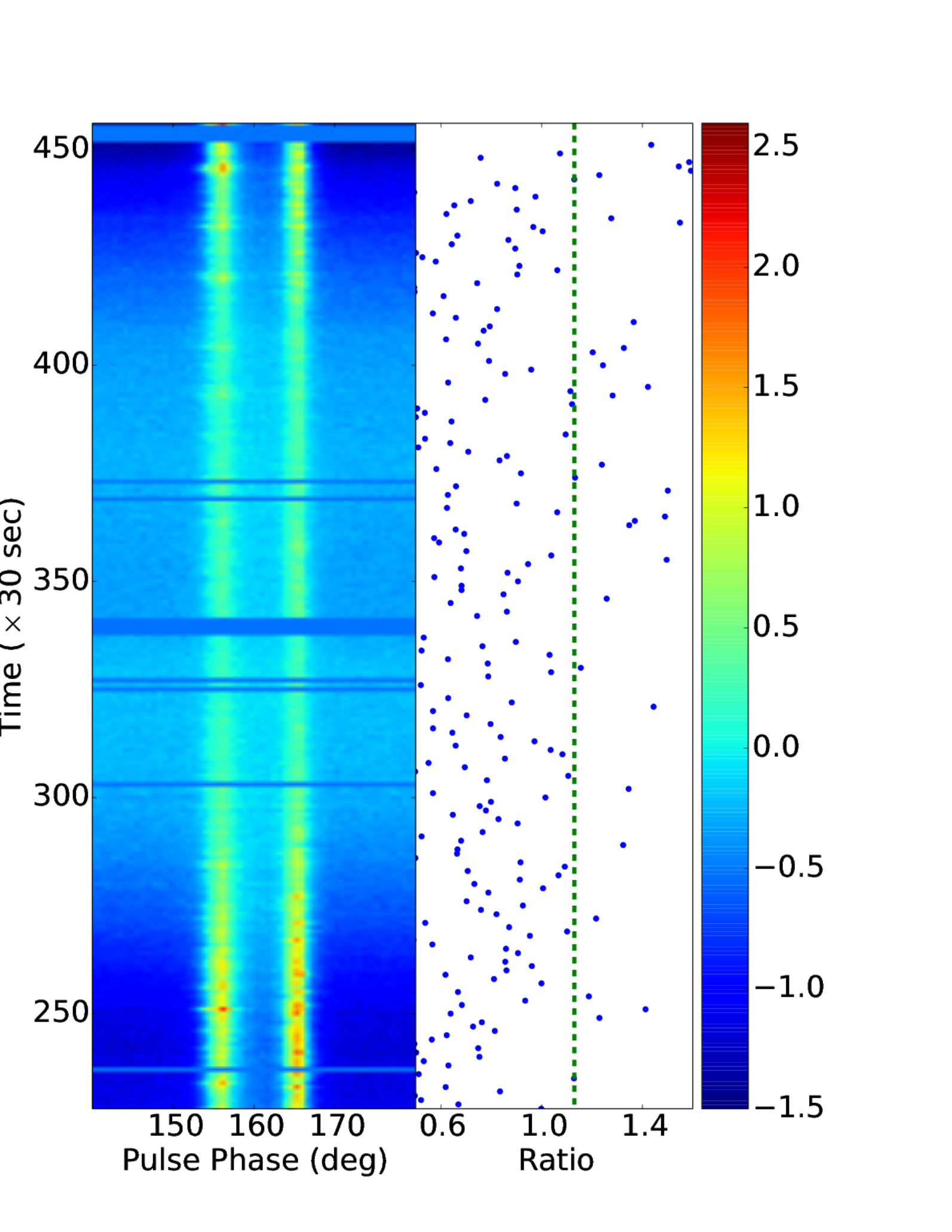}
   \includegraphics[width=8.0cm, angle=0]{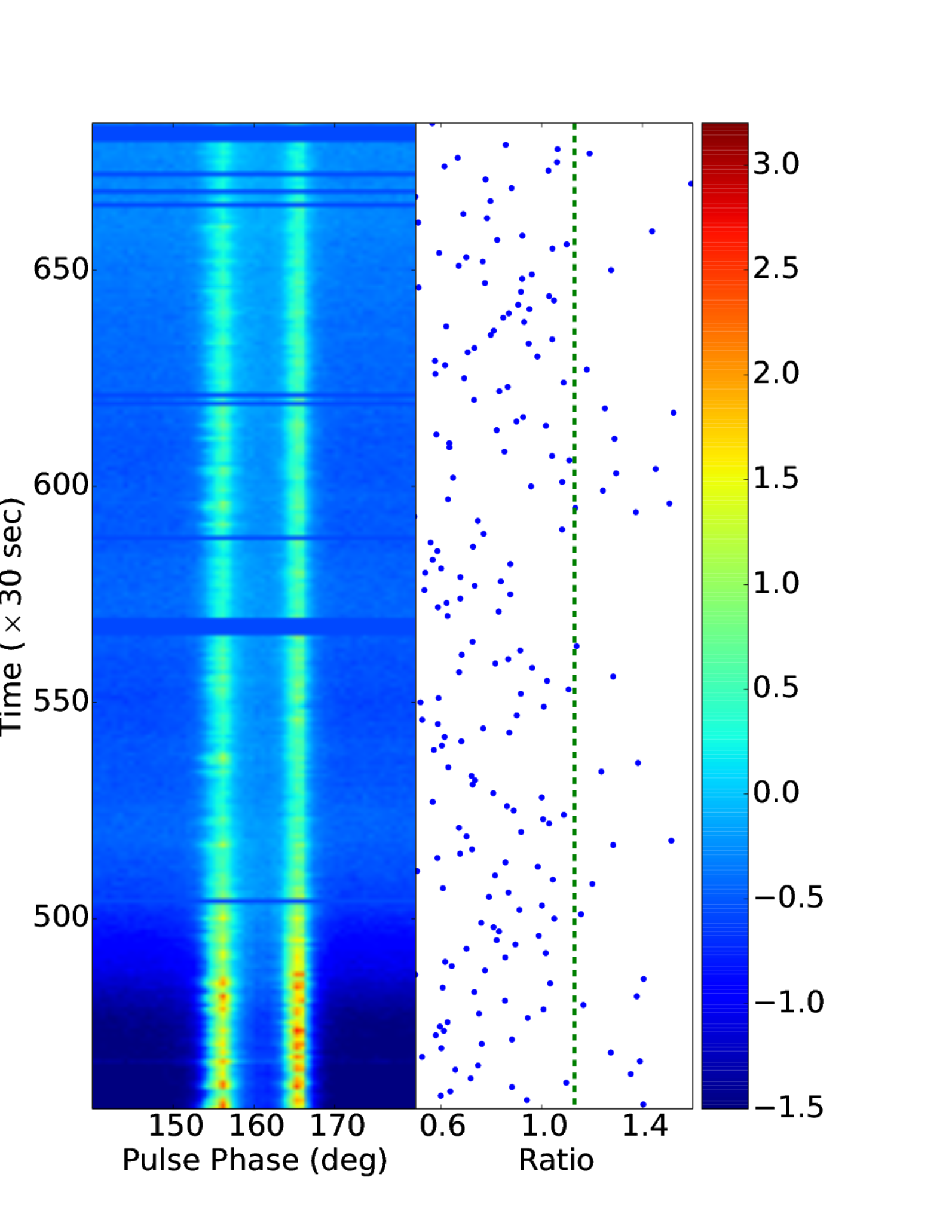}
   \includegraphics[width=8.0cm, angle=0]{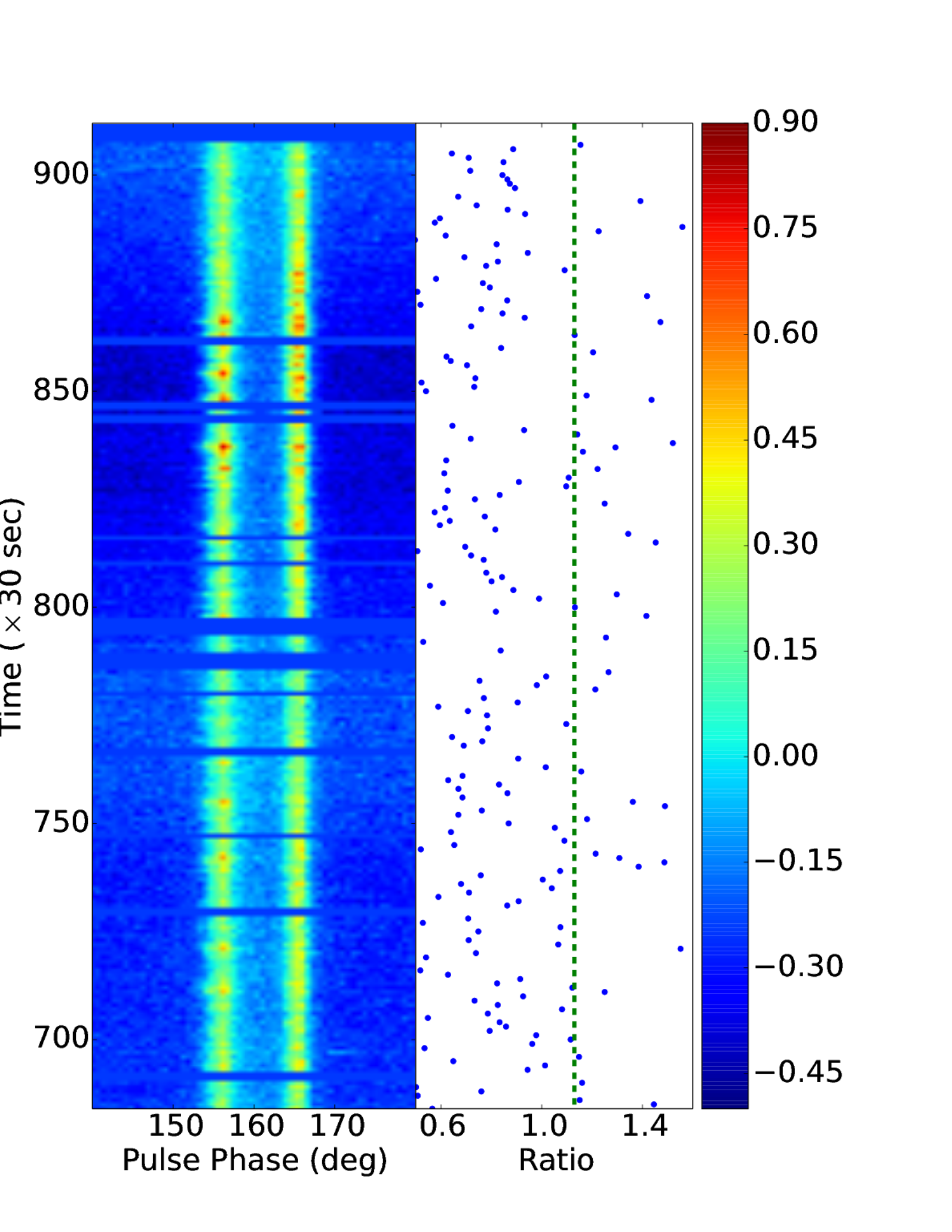}
   \caption{The 912 scintillation-corrected sub-integrations observed at 1556 MHz, each was integrated over 30 seconds. Two-hour blocks are plotted horizontally. The left panels plot the sub-integrations with a colorful density proportional to the received mean flux. Time runs left to right across the pulse window (pulse phase) and bottom to top with sub-integration number. The distribution of the peak intensity ratio between the leading and the trailing components versus the sub-integration number are shown in the right panels. It varies from 0.5 to 1.6. The threshold of normal and abnormal modes are indicated with dashed lines.}
   \label{seq}
\end{figure*}

The whole 8-hour observations with 912 scintillation-corrected sub-integrations, each is averaged over 30 seconds, are shown in Fig. \ref{seq}. The variation of relative peak intensity ($R_I$) between the leading and trailing components are shown in the right panels. It varies frequently in a wide range from 0.5 to 1.6. No strong regularity is found easily by visual inspection. Discrete Fourier Transform (FFT) and auto-correlation were done on the whole 8-hour time sequence of $R_I$ values, and no presence of periodic signals was detected in either methods.

\begin{figure}[h]
   \centering
   \includegraphics[width=8.0cm, angle=0]{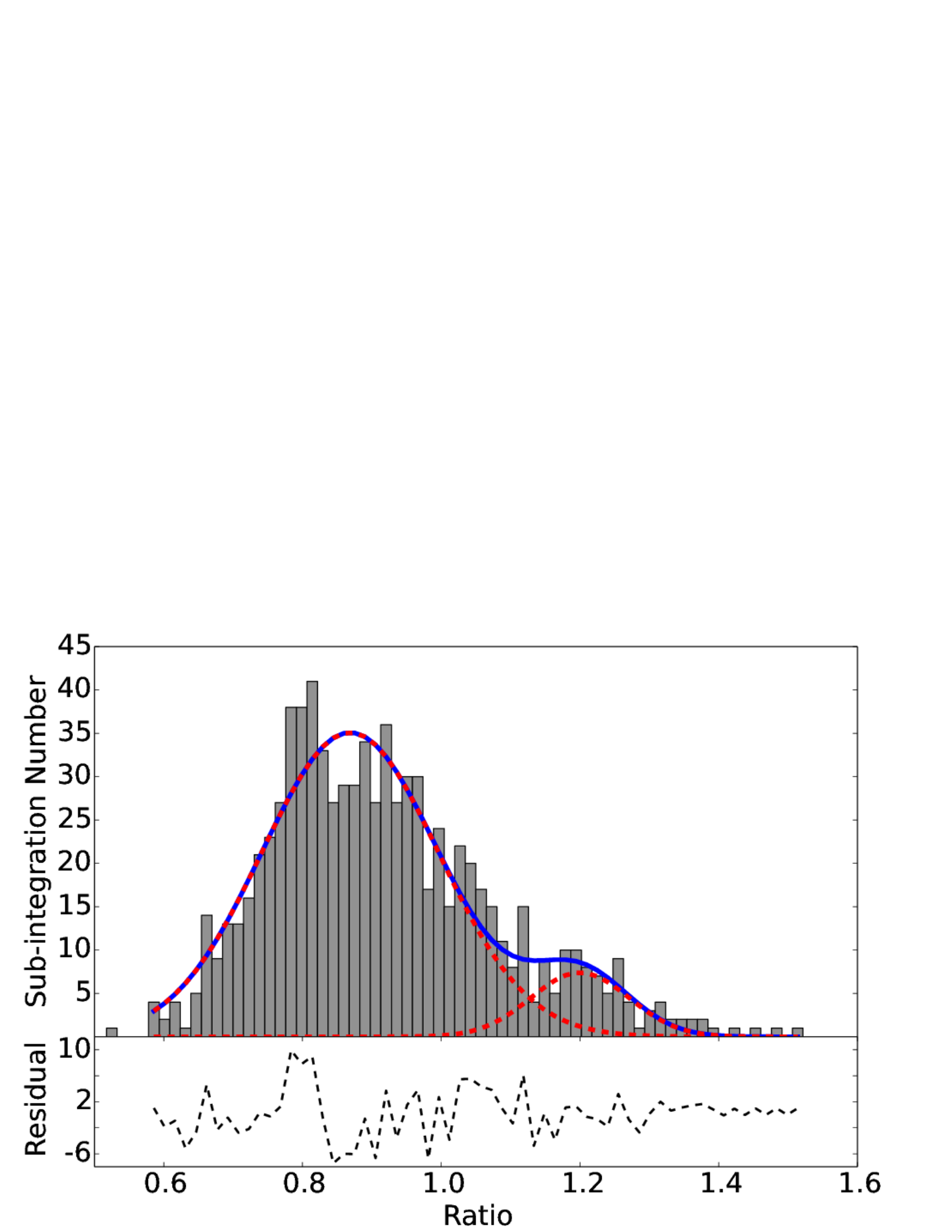}
   \caption{$\rm{R_I}$ distribution versus integration time of 30 seconds by using the data of nearly 8 hours observation. The solid line represents the fitting based on the combination of two Gaussian components (dashed lines). The lower panel shows the fit residuals.}
   \label{histogram}
\end{figure}

The distribution of $R_I$ over nearly 8 hours is shown in Fig. \ref{histogram}, which extends over an extremely wide range. The histogram consists of a broad Gaussian component and a long tail component. As we can see, the best-fit solution gives us a combination of two Gaussian components corresponding to two emission modes. The pulses with $R_I$ in the range of $0.53 \sim 1.13$ are classified as normal mode, the remainder with $R_I$ in the range of $1.13 \sim 1.52$ are classified as abnormal mode. The distribution of the abnormal mode is considerably narrower than that of the normal mode, which may suggest that it is more stable than the normal mode.

\begin{figure*}[h]
    \centering
    \includegraphics[width=12.0cm, angle=0]{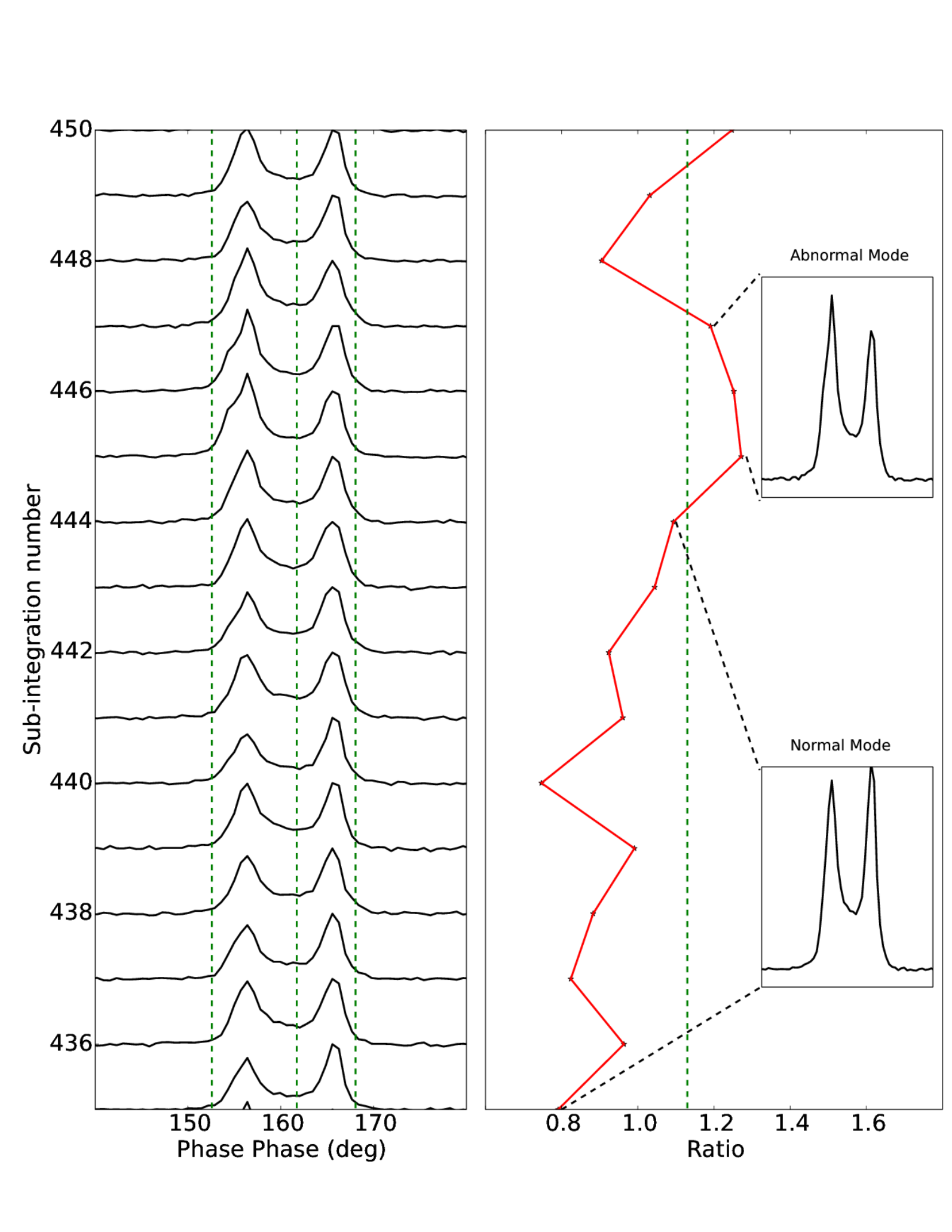}
    \caption{Representative example of a sequence of sub-integrated pulse profiles which contained both the normal and abnormal modes, each averaged over 30 seconds. The 445th to the 447th integrated pulse profiles show switches to the abnormal mode. The dotted lines are the boundaries that divide the profiles into two components, leading and trailing. The right panel shows the distribution of the intensity ratio between the leading and the trailing components versus the sub-integration number. The dashed line shows the threshold of both modes. The averaged pulse profiles of both modes are shown in the insets.}
    \label{demo}
\end{figure*}

Representative example of a sequence of 15 sub-integrated pulse profiles, arbitrarily normalized, is shown in the left panel of Fig. \ref{demo}. The mode switching phenomenon is presented apparently as $R_I$ varies with time, which is shown in the right panel. The integrated pulse profiles from 445 to 447 switch from normal mode to abnormal mode. The averaged pulse profiles of both modes are shown in the insets. Through a careful inspection on the pulse profiles, it is noted that an additional component appears on the leading edge of the first main component for the abnormal pulse profiles.

\begin{figure*}[h]
    \centering
    \includegraphics[width=8.0cm, angle=0]{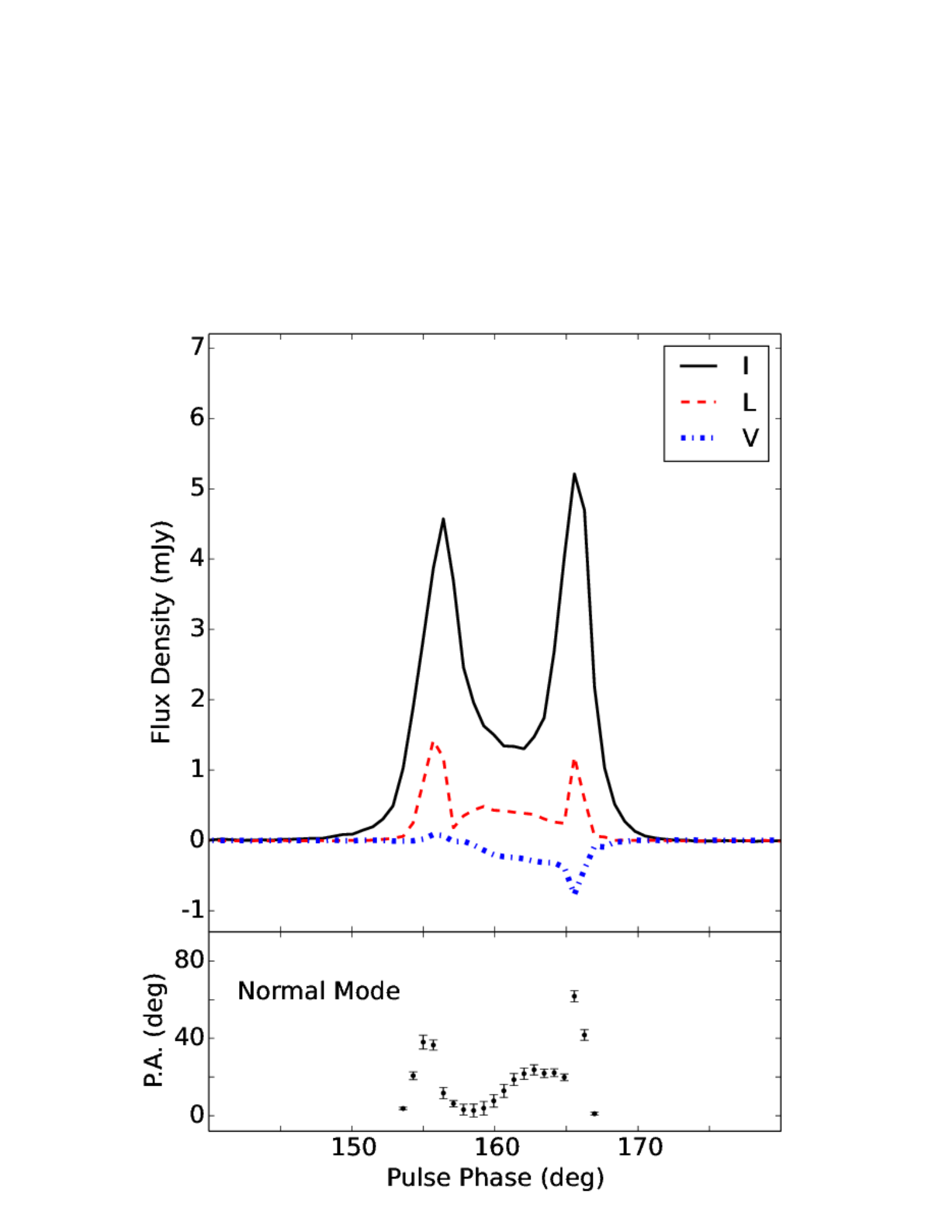}
    \includegraphics[width=8.0cm, angle=0]{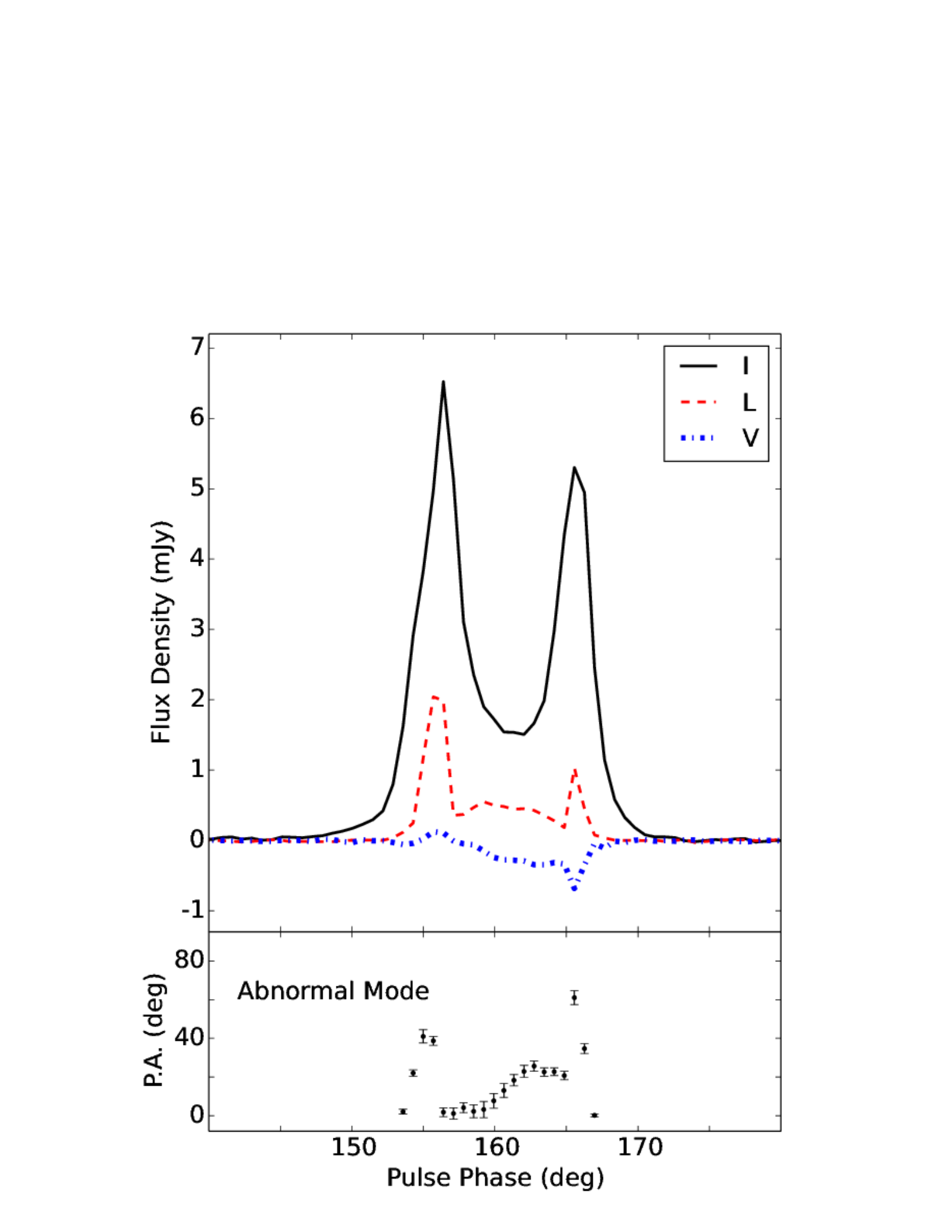}
    \caption{Integrated polarization profiles display for PSR B2020+28 in the normal (left) and abnormal modes (right).}
    \label{Modes_Profiles}
\end{figure*}

The integrated polarization properties of the normal (left panel) and abnormal (right panel) modes are given in Fig. \ref{Modes_Profiles}. The flux density in the abnormal mode is 1.3 times stronger than that of the normal mode.  Note that the polarization profiles are slightly different as well. For the abnormal mode, the linear polarization intensity in the leading component increases by a factor of almost 45\%, the circular polarization intensity decreases by a factor of almost 11\%, compared with those of the normal mode. There are no significant variations in PA between both modes.

\begin{figure*}[h]
    \centering
    \includegraphics[width=8.0cm, angle=0]{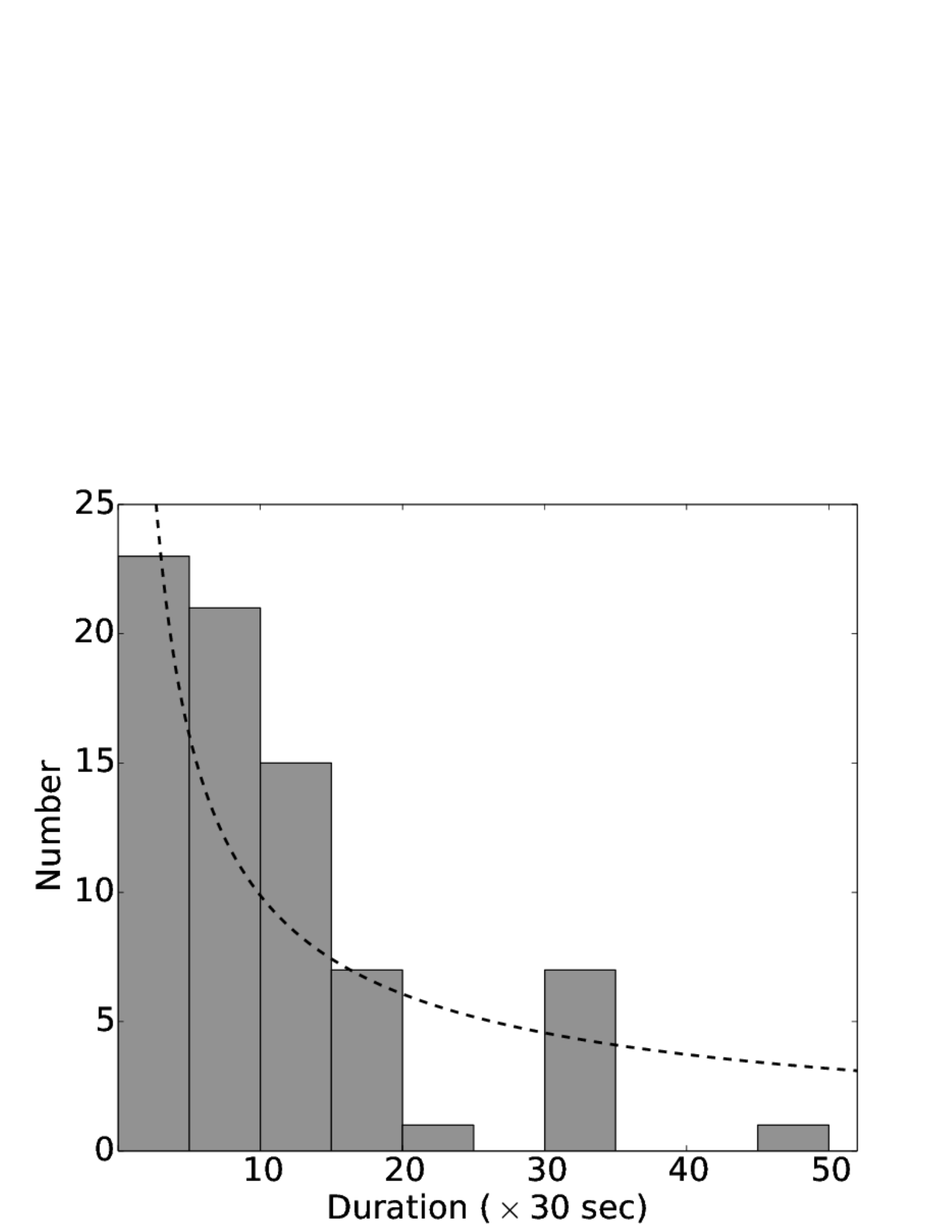}
    \includegraphics[width=8.0cm, angle=0]{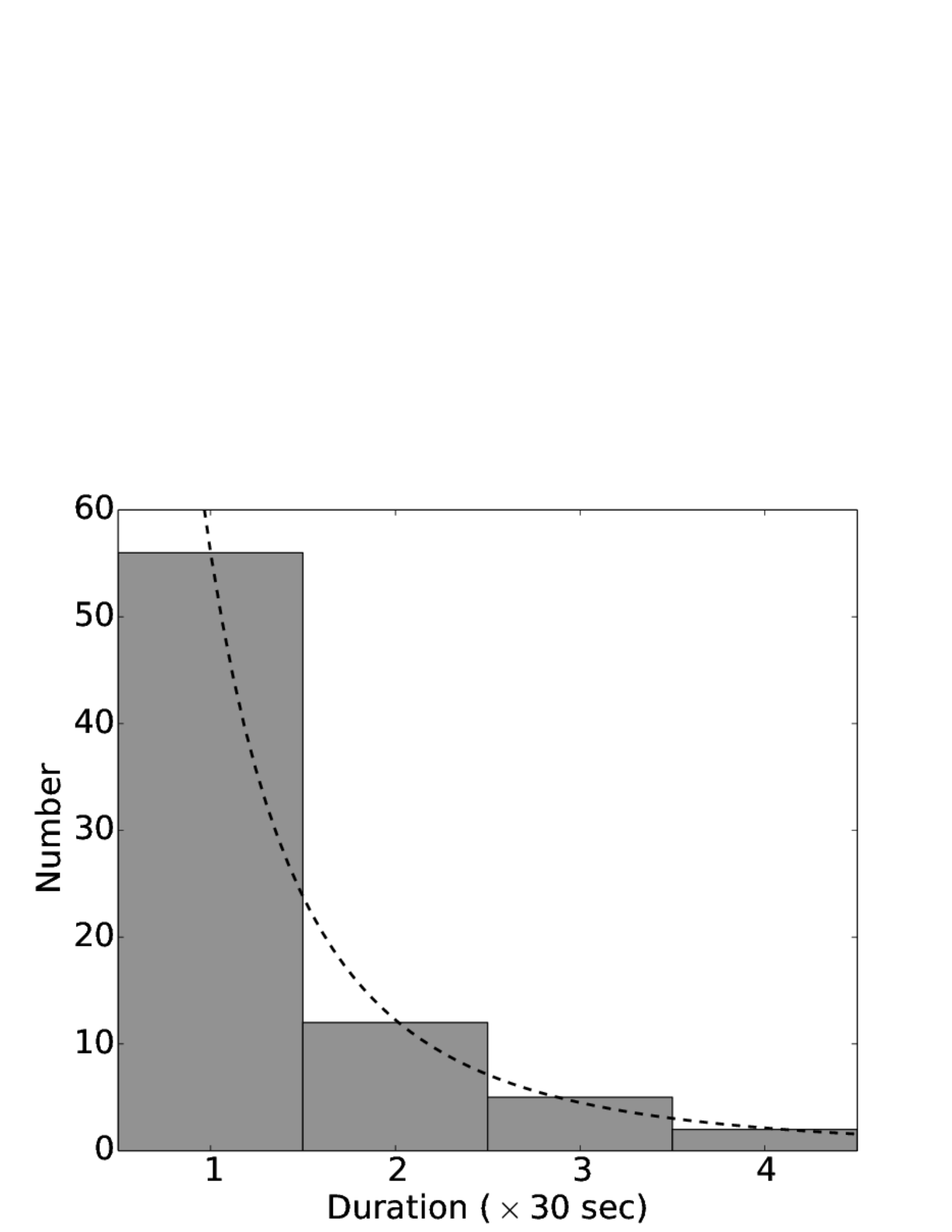}
    \caption{Histograms of the timescales of the normal mode (left) and the abnormal mode (right). The curves stand for the constrained optimal power-law distributions.}
    \label{duration}
\end{figure*}

To further investigate the relative properties of normal and abnormal modes, the durations in each mode are obtained. The histograms for the timescales of the normal (left) and abnormal (right) modes are shown in Fig. \ref{duration}. A curve of best fit was calculated assuming a single power law (dashed line). The exponent, $
\alpha$, derived from the power-law best fits for the normal and abnormal modes are -1.0(1), -2.7(1) respectively. There is a significant difference in the power-law index of the duration distribution between normal and abnormal modes.

\begin{figure*}[htb]
   \centering
   \includegraphics[width=17.0cm, angle=0]{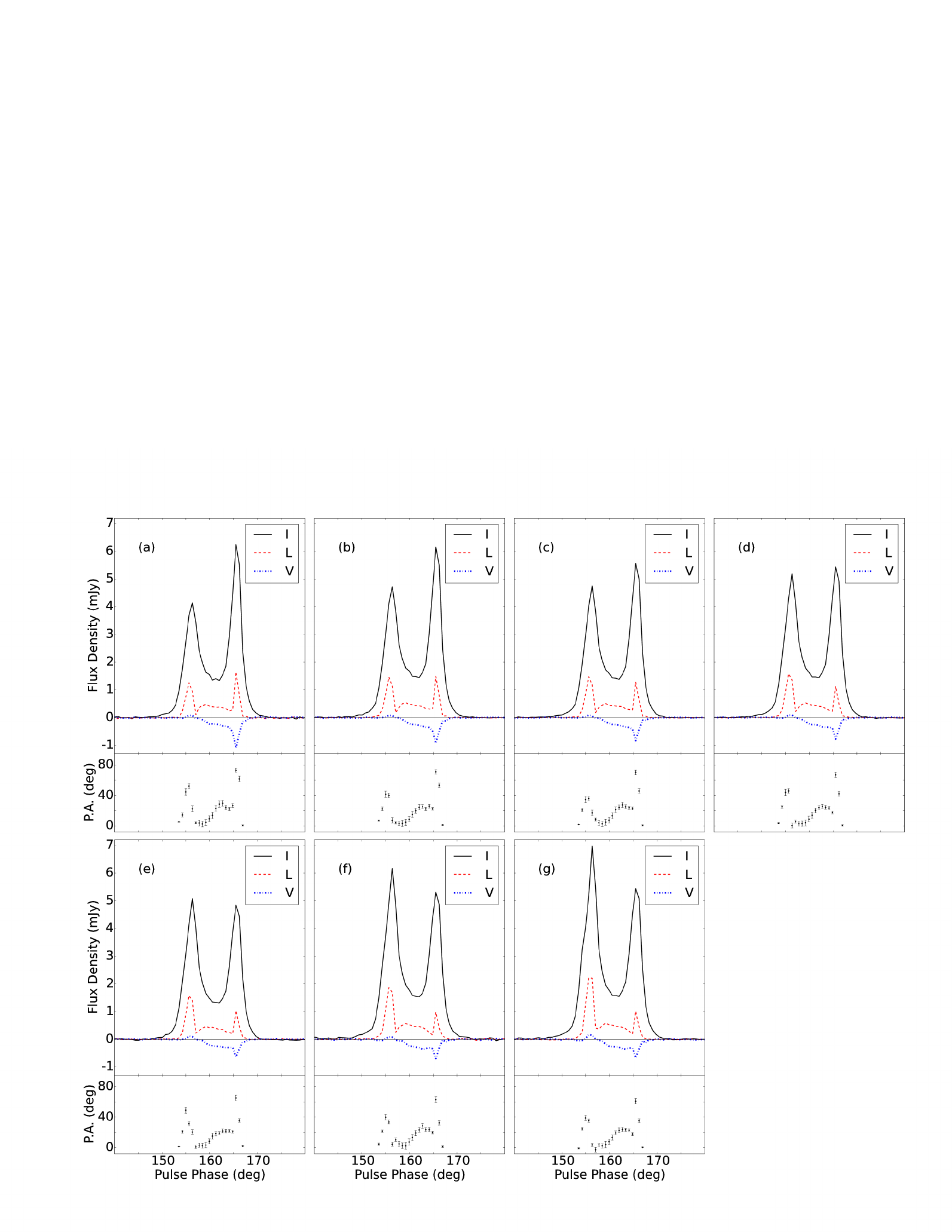}
   \caption{Average Stokes parameter profiles of seven groups of sub-integrations according to the peak intensity ratio ($R_I$) between the leading and the trailing components of the total intensity. The intensity ratios of total and linear polarization have been given in Table~1.}
   \label{waveforms}
\end{figure*}

In order to further to analysis the pulse-to-pulse fluctuation properties. The sub-integrated pulse profiles which are in the same range of $R_I$ are superposed in a group. The details of the division are shown in Table~1. The time  proportions of seven groups are all more than $6$ per cent of the total observation time, which allow high S/N to analysis the polarization properties. The intervals of $R_I$ between two adjacent groups are nearly 0.1. The averaged polarization waveforms of seven groups are shown in Fig. \ref{waveforms}. It is found that the seven groups all have the same pulse widths. The intermediate "saddle" regions in total and linear polarization profiles are relatively broad compared with the widths of the lobes, especially the linear polarization profiles. With the increase of $R_I$, the fractional circular polarization is decreased by a factor of almost 50\%, and a factor of 26\% in the linear polarization is increased. There is no much difference between these seven groups in PA variation. 

\begin{table*}[h]
\begin{center}
\caption[]{The total intensity ratio ($R_I$) and the linear polarization intensity ratio ($R_L$) between the leading and the trailing components of seven different groups. Also present their ratios and errors}\label{Tab1}
\begin{tabular}{clclccccc}
  \hline\noalign{\smallskip}
Group No. & $R_I$ Range & percentage  & $R_I$   & $\sigma$   & $R_L$   & $\sigma$   & $R_L/R_I$    & $\sigma$ \\
  \hline\noalign{\smallskip}
(a) & $<0.7$ & 6.95\%  & 0.6897 & 0.0058   & 0.9558  & 0.0213  & 1.3859    &0.0330\\
(b) & $0.7-0.8$ & 19.12\% & 0.7864 & 0.0029   & 1.1990  & 0.0186  & 1.5246    &0.0244\\
(c) & $0.8-0.9$ & 25.62\% & 0.8768 & 0.0027   & 1.2908  & 0.0192  & 1.4722    &0.0223\\
(d) & $0.9-1.0$ & 22.48\% & 0.9729 & 0.0025   & 1.1649  & 0.0157  & 1.1974   &0.0165 \\
(e) & $1.0-1.1$ & 11.92\% & 1.0711 & 0.0036   & 1.7747  & 0.0297  & 1.6569    &0.0283\\
(f) & $1.1-1.2$ & 7.19 \%  & 1.1710 & 0.0045   & 1.7284  & 0.0393  & 1.476   &0.0340 \\
(g) & $>1.2$ & 6.72\%  & 1.2998 & 0.0041   & 2.1518  & 0.0041  & 1.6555   &0.0263\\
  \noalign{\smallskip}\hline
\end{tabular}
\end{center}
\end{table*}

\begin{figure}[h]
   \centering
   \includegraphics[width=8.0cm, angle=0]{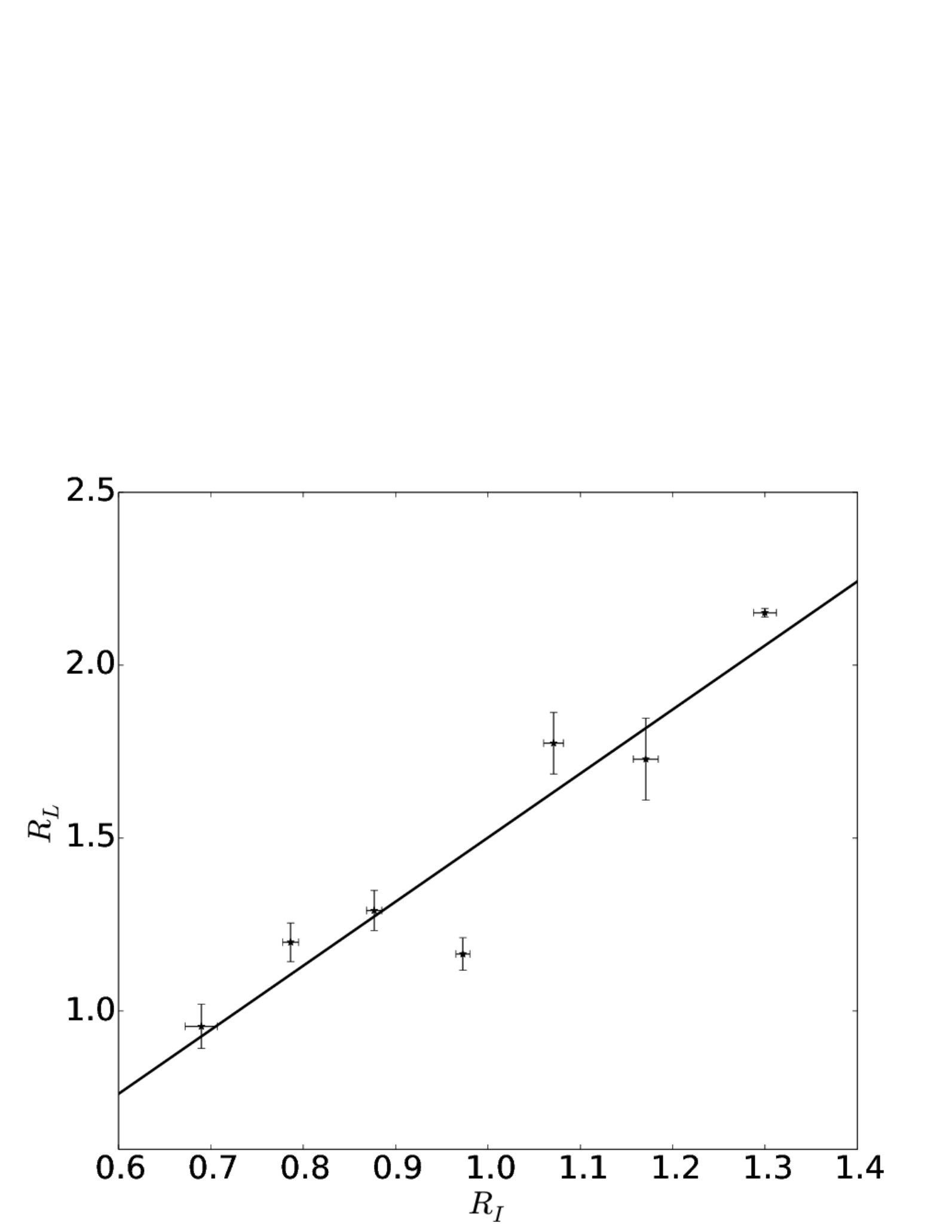}
   \caption{The positive correlation between $R_I$ and $R_L$ after fitting with a straight line. Corresponding error bars denote $3\sigma$ errors.}
   \label{Fit}
\end{figure}

Following to the calculation of $R_I$, the linear polarization intensity ratio ($R_L$) is derived between two peaks in the linear polarization profile. The errors of $R_I$ and $R_L$ are derived from the standard deviation for each group. By means of the error transfer formula, the errors of $R_L/R_I$ are calculated. The intensity ratio ($R_L$) is also changing with that of averaged pulse profiles. And $R_I$ and $R_L$ are positive correlating after fitting with a straight line (see Fig. \ref{Fit}), which indicates that the fractional linear polarization is constant. If this correlation is confirmed in other pulsars, $R_L$ may be also used as an indicator to identify the mode switching phenomenon and to figure out emission mechanism more than $R_I$.

\section{Discussion and conclusions}

We report new results on the emission properties of pulsar B2020+28 based on detailed analysis of 8-hour observations at a center frequency 1556~MHz using the Nanshan 25-m radio telescope. A total of 76 mode switching events are detected. It spends 89\% in the normal mode and 11\% in the abnormal mode.

The major difference between the normal and abnormal modes is the intensity ratio between the leading and trailing components and the length of moding timescales. The variation of $R_I$ has no strong regularity, and the mode switching phenomenon seems a random process. The distribution of $R_I$ for the abnormal mode is narrower than that of the normal mode by a factor of 65\%, which may indicate that the abnormal mode is more stable than the normal mode. The intrinsic timescale distributions, constrained for this pulsar for the first time, provide valuable information to understand the physics of mode switching phenomenon. The durations of abnormal mode are extremely short, which are less than 250 pulse periods. They may indicate that the timescale for the abnormal mode to get stable is shorter than that for the normal mode. The short durations of both modes are very common and long durations are much less, which may be represented by an elementary Markov process \citep{Cordes+2013}.

The strong linear polarization is an outstanding characteristics of pulsars, which is usually associated with cone emission around magnetic field lines \citep{Gedalin+Dzigan+2005}. The stronger linear polarization in the leading and trailing components than that in the intermediate "saddle" region reveals the double-lobed structure in the total intensity is from the conal emissions. The circular polarization often accompanies core emission and is generally the strongest in the central or 'core' regions of a profile. The sense reversal often occurs near the middle of the profile \citep{Rankin+1983}. But for PSR B2020+28, the circular polarization changes senses at the leading component and reaches peak value at the trailing component.

The mechanisms of mode switching phenomenon are proposed by many authors. Pulsar switches between different magnetospheric states are likely to be caused by changes of  particle current flow in the pulsar magnetosphere \citep{Lyne+etal+1971,Bartel+etal+1982}. In some cases, the pulse profile changes are also correlated with large changes in spin-down rates. \citep{Lyne+etal+2010}. Changes between two distinct emission states in PSR J0742$-$2822 is correlated with the changes in the derivative of the pulse frequency \citep{Keith+etal+2013}. The alteration of the temperature of pulsar surface could trigger different sparking modes in the inner vacuum gap, thus results in mode switching phenomenon \citep{Zhang+etal+1997}. The switches in the magnetosphere geometry or/and redistribution of the currents flowing in the magnetosphere change the pulsar emission beam and its orientation with respect to the line of sight and hence lead to the mode switching phenomenon \citep{Timokhin+2010}. For PSR B2020+28, the frequent switching between normal and abnormal modes suggests that the oscillations between two different magnetospheric states are rapid.

Normally, the pulse-to-pulse variability are hard to identify in sources, because summing over many pulses is required to achieve sufficient S/N. To distinguish mode switch and pulse-to-pulse fluctuations in pulse shapes, fitting to the distribution of $R_I$ with two Gaussian components has smaller residuals than that with one. 
Furthermore, the whole data set are examined carefully, approximately third of pulse profiles show an additional leading component in the intervals of abnormal mode. And this additional component is also detected in several normal pulse profiles. As shown in Fig. \ref{duration}, the duration of abnormal mode is relatively short, the detected abnormal pulse profiles may be contaminated by some normal pulses, so the additional weak leading emission becomes blurred. Therefore, the association between the additional component with intervals of abnormal mode is expected to exist in PSR B2020+28.

The relative pulse-to-pulse total intensity fluctuation is identified in variable linear and circular polarization. The evolution of fractional polarization with the increase of $R_I$ was also identified. The stable PA variations of different groups imply that the geometry of the pulsar emission beam remains constant, while the emission strength varies in different emission regions, as a result of the intensity variations of total and linear polarization. It is noted that most points deviate from the straight line shown in Fig. \ref{Fit}. Such deviation maybe due to the error in the systematic polarization calibration, or perhaps caused by the intrinsic variation.

Mode switching is expected to occur over a broad-range of wavelengths \citep{Sobey+etal+2015}, even up to X-rays. Therefore, we would require multi-frequency simultaneous single pulse observations to better understand these emission characteristics, which may lead to a better understanding of the magnetospheric emission mechanism.

\acknowledgments
We are grateful to the referee for valuable suggestions. This work was supported by National Basic Research Program of China grants 973 Programs  2015CB857100 and 2012CB82180, the Pilot-B project grant XDB09010203, and the West Light Foundation of Chinese Academy of Sciences (WLFC) No.XBBS201422. WMY is supported by NSFC (11203063, 11273051) and \\WLFC (XBBS201123). JPY is supported by NSFC 2012CB821801 and 11173041. We thank members of the Pulsar Group at Xinjiang Astronomical Observatory for helpful discussions.

%\bibliographystyle{spr-mp-nameyear-cnd}  
%\bibliographystyle{mn2e}
%\bibliography{2020}

\end{document}